# A Logic Simplification Approach for Very Large Scale Crosstalk Circuit Designs

Md Arif Iqbal[1], Naveen Kumar Macha, Bhavana Tejaswini Repalle, Mostafizur Rahman[2]
Computer Science Electrical Engineering, University of Missouri-Kansas City, MO, USA
mibn8@mail.umkc.edu[1], rahmanmo@umkc.edu[2]

*Abstract*— Crosstalk computing, involving engineered interference between nanoscale metal lines, offers a fresh perspective to scaling through co-existence with CMOS. Through capacitive manipulations and innovative circuit style, not only primitive gates can be implemented, but custom logic cells such as an Adder, Subtractor can be implemented with huge gains. Our simulations show over 5x density and 2x power benefits over CMOS custom designs at 16nm [1]. This paper introduces the Crosstalk circuit style and a key method for large-scale circuit synthesis utilizing existing EDA tool flow. We propose to manipulate the CMOS synthesis flow by adding two extra steps: conversion of the gate-level netlist to Crosstalk implementation friendly netlist through logic simplification and Crosstalk gate mapping, and the inclusion of custom cell libraries for automated placement and layout. Our logic simplification approach first converts Cadence generated structured netlist to Boolean expressions and then uses the majority synthesis tool to obtain majority functions, which is further used to simplify functions for Crosstalk friendly implementations. We compare our approach of logic simplification to that of CMOS and majority logic-based approaches. Crosstalk circuits share some similarities to majority synthesis that are typically applied to Quantum Cellular Automata technology. However, our investigation shows that by closely following Crosstalk's core circuit styles, most benefits can be achieved. In the best case, our approach shows 36% density improvements over majority synthesis for MCNC benchmark.

*Keywords*— *Crosstalk Computing, Capacitive Coupling, Crosstalk Logic, Majority Network, Logic Synthesis*

## I. INTRODUCTION

As traditional way of CMOS scaling becomes difficult, Crosstalk computing provides an alternative solution while leveraging CMOS devices and interconnect technologies [1]-[5]. In Crosstalk circuits, computation is realized by embracing the increasing crosstalk signal interference at advancing technology nodes and astutely engineering it to a logic principle. For operation, the transition of signals on input metal lines called as aggressor nets, induce a resultant summation charge on output metal line, called as victim net, through capacitive couplings. This induced signal serves as an intermediate signal to control thresholding devices like an inverter to get the desired logic output.

All the elementary gates, as well as many multi-level logic functions, can be implemented by a single Crosstalk gate. To implement a multi-level logic function, two different circuit implementation styles are followed which are homogeneous and heterogeneous. In homogenous circuits, the coupling capacitance between input and output nets are equal, whereas in heterogeneous, the capacitances are unequal. Crosstalk circuits use these homogeneous and heterogeneous cells as primitive cells along with other gates like AND/OR. Due to the innovations in circuit style and physical principle of computing, the traditional synthesis flow for large circuits is not directly applicable.

Majority logic, where the summation of signals determines logic output through thresholding function, can resemble some of the Crosstalk's logic principles. However, existing majority synthesis approaches in literature mostly concentrate on Quantum Cellular Automata technology whose primitive cells are only inverter and majority gates [6]-[8]. Though some benefits can be obtained by using majority synthesis methods, fundamentally, obtaining simplified expressions for Crosstalk circuits require a different approach that utilizes fabric's native functionalities.

We propose a Crosstalk implementation friendly logic simplification approach that takes advantage of both CMOS and majority synthesis methods for simplified Boolean expressions. First, we take an arbitrary the network in Verilog form and use Cadence RTL Compiler [9] to generate a netlist of the network with logic constraints (e.g., limit the tool to use NAND/NOR, AOI, OAI gates only) to benefit from Crosstalk implementations. Then this netlist is converted to Boolean expressions and fed to the SIS [10] tool to obtain 3-input Boolean expressions. These expressions are then used in our logic simplifier tool to iteratively get Crosstalk friendly expressions Our results show that for three different circuits' cm85, mux and pcle from MCNC benchmarks [11], there are 11%, 27%, and 32% transistor count reduction compared majority synthesis approach and 58%, 62%, and 24% transistor count reduction compared CMOS based approach respectively.

The rest of the paper is organized as follows: Section.II describes the fundamentals of Crosstalk computing (CT) and implementation of logic gates. Section.III presents the overview of logic simplification methodology. Section.IV compares and benchmarks proposed simplification methods with majority based synthesis and CMOS based synthesis methods and Finally, Section.V presents the conclusion.

## II. CROSSTALK CIRCUIT STYLE & POTENTIALS FOR DENSITY, POWER AND PERFORMANCE GAINS

Interference that is observed between metal lines in advance technology, we want to control it such that we can engineer to obtain logic. Our engineering takes place on the organization of the metal lines and weight or value of capacitances which are put between them to get the logic functionality. For example, in Fig.1 we show a NAND get can be achieved by the interference of two signals Ag1 and Ag2. Here, both capacitances have the same value, $C_{ND}$. By changing this capacitance to a different value, we can ensure that when one of the signals transition, we

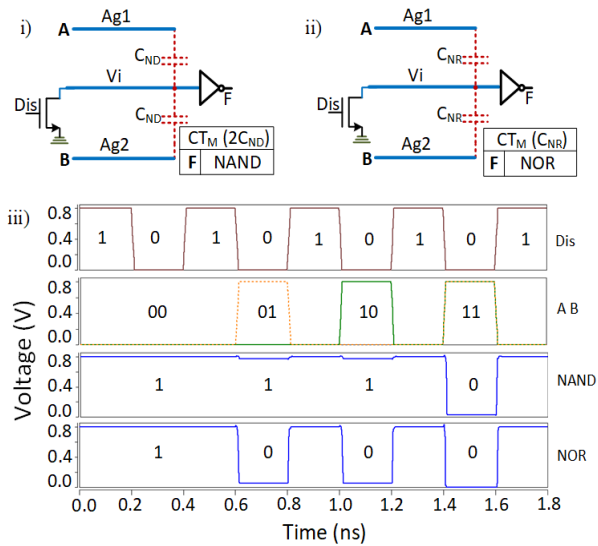

Fig.1 Fundamental Logic Gates: i) NAND ii) NOR have an output in the victim node which results in OR behavior and with the inverter, we achieve full swing.

Figs 1(i&ii) shows the implementation of primitive Crosstalk cells NAND and NOR with different capacitances ($C_{ND}$ & $C_{NR}$) however between the inputs, capacitances are kept same. If we alter the capacitances between one input and another input, we can achieve different heterogeneous functionalities termed as heterogeneous Crosstalk logic. One example of this is AOI21 (AND-OR-Inverter), i.e., $F = (AB+C)'$ which is shown in Fig 2 (i&ii).

Noticeably, when we have homogeneous functionality with multiple inputs, this has some similarity to Majority logic where majority threshold functions are generally used to obtain max/min functions. An example of this can be carry logic $F = MAJ_3(A,B,C) = AB + BC + C$ shown in Fig 2(iii & iv). However, the key difference is that we can achieve not only Boolean logic gates (NAND, NOR) and Majority logic but also heterogeneous logic. For us, the flexibility is much more than just majority gates or that of CMOS which provides more opportunities to compress logic using Crosstalk logic cells.

For large-scale circuits, logic cascading and maintaining signal integrity is a critical issue. In this regard, the crosstalk computing approach provides opportunities as well as challenges. Since utilizing crosstalk we can implement both fundamental logic gates and reduce complex combinational logic blocks, any logic function can be implemented. The logic functions that require hierarchical implementation will be implemented by cascading outputs through the coupling. In this regard, we can use a constructive-destructive topology that is if a non-inverted gate is implemented first, we can cascade the output to an inverting gate or vice-versa. While cascading outputs at several levels, maintaining signal integrity becomes a challenge, since with each stage of coupling the induced voltage in the next level reduces compared to previous. We resolved this issue in different ways by placing buffers or by using a Pass-Gate solution, where, the inverting and non-inverting gate interfaces are connected through a transmission gate which is controlled by clock cycles [2]. The other solution is by using a different set of Crosstalk logic gates which operate on falling edge transition also. Thus, a fully working large-scale compact circuits, with reduced size, improved performance and power can be achieved using Crosstalk logic style.

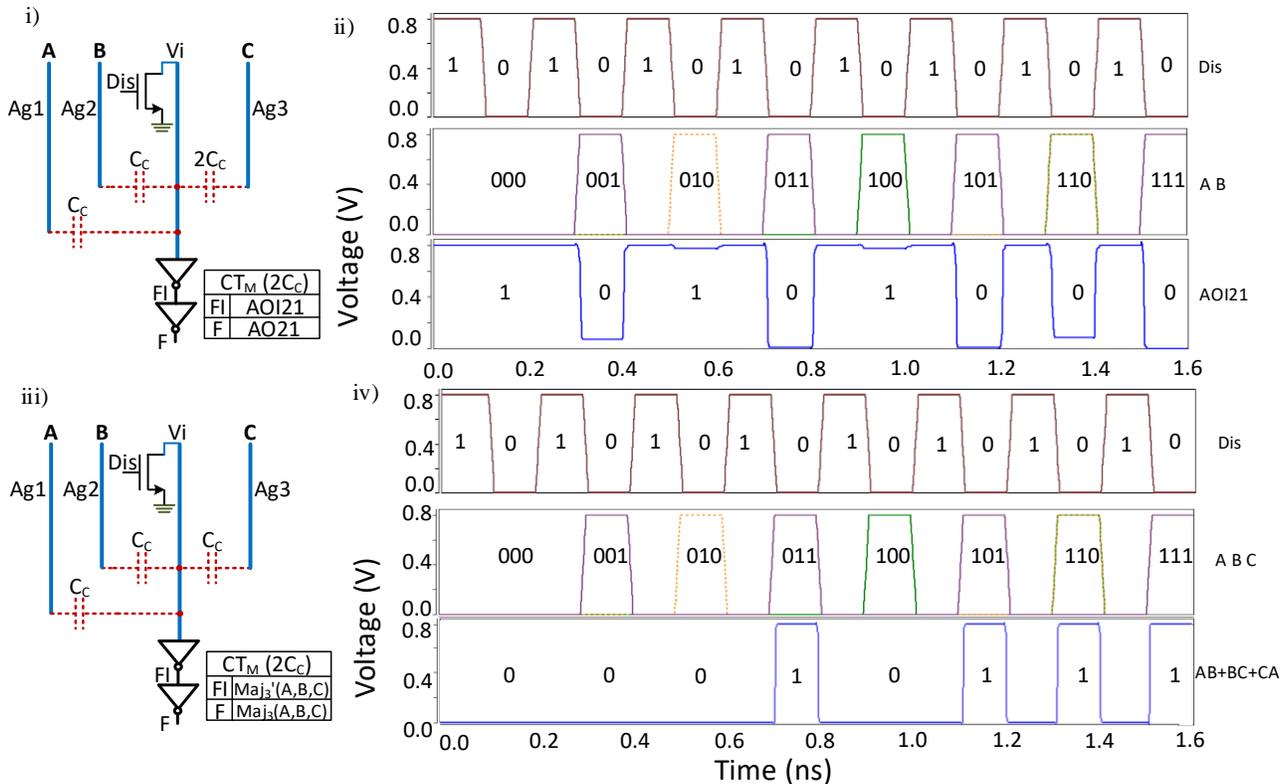

Fig. 2. CT Heterogeneous and Homogenous gates, i) AOI21 (AND-OR-Inverter), ii) Simulation results of AOI21, iii) $f=AB+BC+CA$, iv) Simulation results of function $f$.

### III. OVERVIEW OF THE PROPOSED LOGIC SIMPLIFICATION METHODOLOGY

In this section, we introduce our simplification approach for Crosstalk circuit friendly expression and detail implementation steps. We take advantage of the compressibility feature that Crosstalk presents through custom logic, CMOS logic, Majority logic and explain in our approach that how we can combine all of them to obtain the best result.

Fig. 3(i&ii) gives a flow diagram of crosstalk logic synthesis methodology. Our process starts with having Cadence RTL Compiler that generates a netlist from Verilog code with constrains such as that it has to use gates like NAND/NOR/AOI which are Crosstalk friendly. It is noticeable though we cannot constrain the tool to use majority gates (AB+BC+CA) or other heterogeneous logics that are especially suitable for Crosstalk. Because of this, once we obtain netlist from Cadence tool then we convert it back to Boolean expressions and feed it again through SIS tool such that the SIS tool already works on an optimized Boolean expression and further tries to simplify it in terms of majority gates (fig. 3(i)). Since the expression already has majority expressions and some custom expressions which can be implemented using universal gates like NAND/NOR gate but we look for further opportunities for simplification as given in fig. 3(ii) to get expressions for heterogeneous logic. If the heterogeneous logics cannot be found we use Crosstalk NAND/NOR gate and complete Boolean expression. Finally, we obtain a final expression that can be converted into structural netlist and that structural netlist can be used in conjunction with cell libraries to obtain full layout and parametric results like area, power, and performance.

Fig. 4 represents the pseudo-algorithm of our simplification approach where we check for each function of the network to be simplified as crosstalk friendly expression. Variables that are used in the algorithm are defined as follows:

| | |
|---|---|
| $f_1, f_2, f_3$ | A function in network N |
| S | Set of Crosstalk homogeneous & heterogeneous function |
| $f_n$ | Fannin to network N |
| S' | Inverted Crosstalk homogeneous & heterogeneous function |
| $l_i$ | The $i^{th}$ literal in the expression for function $f$ |
| $p_j$ | The $j^{th}$ product term for function $f$ |
| $n_l$ | No. of literals in function $f$ |
| T | No. of the transistor in the function |
| I | No. of the inverter in the function |
| $f_{dm}$ | Function $f$ after applying De Morgan's Law |

The corresponding pseudo algorithm takes in preprocessed and decomposed network as input and returns a more simplified network that is crosstalk friendly.

After preprocessing and decomposing, each function $f$ of the network N is checked to determine if it is in homogeneous or heterogeneous crosstalk form. If it is, we proceed to simplify the next function. Otherwise, as shown in Fig. 3(ii), we check to see if there exist more than two literals in the function. If there exist only two literals in the function, we check for transistor count. First, we calculate no. of transistors need to implement function $f$. Then, we take an inverted function ($f_x$) of $f$ and calculate a number of transistors required. If the transistor count for $f_x$ is lower than the original function $f$, we update the function with $f_x$. For example, consider a function $f=a+b'$. Crosstalk mapping

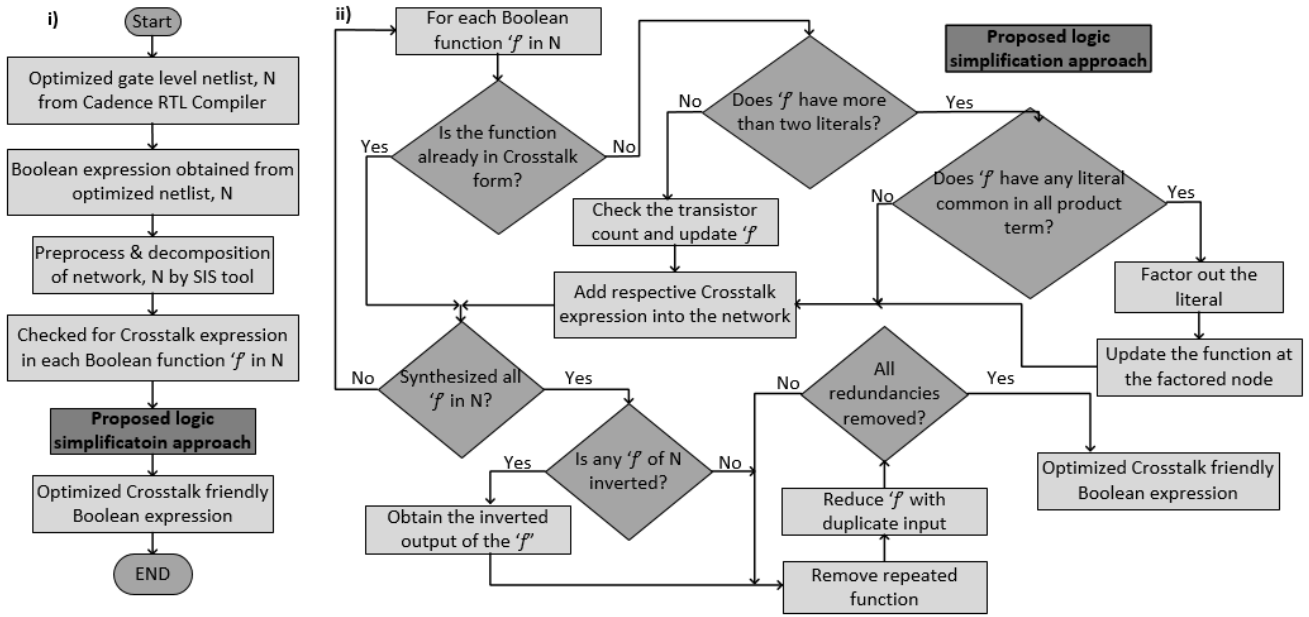

Fig. 3. Overview of proposed logic simplification. i) Top-level simplification approach, ii) Detail steps of proposed logic

```
Input: Optimized Network N
Output: Crosstalk expression corresponding to N
begin
1      Convert the netlist to Boolean expression
2      Preprocess and decompose network N by SIS
3      for each f in N do
4          if f ∉ S then
5              if n_l > 2 then
6                  if ∃l_i so that ∀j, l_i ∈ p_j then
                       f_1 = f|_{li = 1}
                       f = l_i f_1
7                  else
                       Add Crosstalk expression to the
   f
8              else
                   //check the transistor count
                   Count f_old = T+2*I
                   Apply De Morgan's Law to function f
                   f_x = f_dm'
                   Count f_new = T+2*I
9                  If Count f_new < Count f_old then
                       f = f_x
10                 else
                       Keep the original f and add
                       Crosstalk expression
11         else
               Add Crosstalk expression to the f
12     if f_n in N such that f_n = f' do
           f_n = S' where S' is the inverted form of S
13     else
           break
14     Do redundancy check
15 end
```

Fig 4. Pseudo algorithm for Crosstalk logic synthesis

would require seven transistors including an inverter for literal b to map the function f. However, inverted function $f' = f_x = (a'b)'$ would require only five transistors. If there are more than two literals present in the function, we look for any common literal that is present in all the product terms of the function. If a common literal exists, we factor this literal out and mapped with heterogeneous crosstalk circuits. Consider function $f = bc + ca$. If we are to map crosstalk gates directly, it would take three crosstalk gates whereas if we factor out the common literal c from both product terms, function f therefore can be presented as $f = cf_1$, where $f_1 = (b+a)$, thus requiring only one crosstalk gate. If there are no common literals, we check whether all the functions are synthesized or not. After simplifying all the functions in the network N, we further investigate if there exist any function that is in inverted form. If so, by using Crosstalk fabric inherent feature, we can save any additional inverter, required for making function f inverted. The final process is to remove all redundancies, if exist, otherwise terminate. For redundancy removal, we follow the procedure explained in [8].

Next, we present two networks of Boolean expressions to explain the flowchart. First, Boolean expression is obtained from 4-bit ALU and the second one is the network for the 2-bit multiplier. We represent the Crosstalk functions by denoting as function $X_{gate}(a,b,c)$ where a,b,c are the sub functions and subscript gate defines what type of logic the function will implement.

**First example:**

**Step 1:** Boolean expression obtained from 4-bit ALU.

$((((A_1A_2+A_1B_2')+A_2(B_2'+B_1'))+B_1'B_2')A_3+(((A_1A_2+A_1B_2')+A_2(B_2'+B_1'))+B_1'B_2')B_3'+(A_3+(B_0'B_3'))')+A_3B_2B_3'$

**Step 2:** By using SIS [10] tool for preprocess and decompose, we obtain the following expression

$$N = f_2 + f_3 + f_5$$
$$f_1 = A_1 + B_1' \quad (1)$$
$$f_2 = A_3B_0'B_3' \quad (2)$$
$$f_3 = A_3B_2B_3' \quad (3)$$
$$f_4 = f_1A_2 + f_1B_2' + A_2B_2' \quad (4)$$
$$f_5 = f_4A_3 + f_4B_3' \quad (5)$$

**Step 3:** For each function of network N, presented in equation (1)-(5), we have to check if the function is already in Crosstalk homogeneous or heterogeneous form.

- The first function, $f_1$ is neither in homogeneous nor in heterogeneous form. Next, we found that it has only two literals. Then, we checked for fewer transistor count which we got after applying De Morgan's law and then taking inverter of the function f. Therefore, the updated function would be $f_1 = (A_1'B_1)'$. Since there are still three other functions to be simplified, we proceed to the next function, $f_2$.
- Function $f_2$ is directly in crosstalk homogeneous form $X_{and}(A, B, C)$. We proceed to simplify the next function.
- Function $f_3$ is also directly in crosstalk homogeneous form $X_{and}(A, B, C)$. Therefore, we update the function with crosstalk homogeneous expression and check if there is any other function left to be simplified.
- Function $f_4$ is in crosstalk homogeneous form $X_{homo}(ab+bc+ca)$ too, so, we update the function with crosstalk homogeneous gate.
- Function $f_5$ is neither in homogeneous nor in heterogeneous form. Next, we checked to see if the function has any common literals. We found that $f_4$ is the common literal in both of the product terms in function $f_5$. Therefore, we factor out the common term and update the function as $f_5 = (A3+B3')f_4$ which is in heterogeneous crosstalk circuit $X_{hetero}((A+B)C)$ form. Next, we proceed to simplify other functions.
- From equation (1), we can see that both function $f_2$ and function $f_3$ have common literals $A_3B_3'$ between them which we can factor out and get the expression as $A_3B_3'(B_0'+B_2)+f_5$. $A_3B_3'$ term can be obtained by Crosstalk and gate which we can AND with $(B_0'+B_2)$ to get Crosstalk heterogeneous form.

**Step 4:** Update the node function for inverted output. Since there is no other function to simplify, we check if there exist any function in inverted form. If so, we can avoid additional inverter by using Crosstalk fabric feature, which can apply for the fan-in $f_3'$.

Table I
Comparison of density reduction for different Boolean Network

| Standard Function | I/O | CMOS | | Synthesis using existing method | | Synthesis using proposed method | | R% w.r.t CMOS | | R% w.r.t existing method | |
|---|---|---|---|---|---|---|---|---|---|---|---|
| | | Transistor Count | Gate Count | Transistor Count | Gate Count | Transistor Count | Gate Count | Transistor Count | Gate Count | Transistor Count | Gate Count |
| F=ab+bc+a'b'c' | 3/1 | 30 | 7 | 20 | 6 [7] | 13 | 3 | 56% | 57% | 35% | 50% |
| F=d(c+(b'+a)') | 4/1 | 18 | 4 | 25 | 6 [8] | 12 | 3 | 33% | 25% | 52% | 50% |
| Example1 | 7/1 | 94 | 21 | 62 | 16 [8] | 44 | 11 | 53% | 48% | 29% | 31% |
| **Arithmetic Block** | | | | | | | | | | | |
| Full Adder | 3/2 | 18 | 9 | 17 | 4 [8] | 10 | 2 | 44% | 77% | 41% | 50% |
| 2-bit Multiplier | 4/4 | 56 | 15 | 67 | 17 [8] | 43 | 13 | 23% | 13% | 36% | 23% |
| **MCNC Benchmark** | | | | | | | | | | | |
| cm85a | 11/3 | 264 | 64 | 125 | 31 [8] | 111 | 27 | 58% | 58% | 11% | 13% |
| mux | 21/1 | 404 | 72 | 209 | 49 [8] | 152 | 37 | 62% | 49% | 27% | 12% |
| pcle | 19/9 | 246 | 56 | 276 | 66 [8] | 186 | 42 | 24% | 25% | 32% | 36% |

**Step 5:** Check for redundant functions and also redundant input to single functions. We have checked and found no redundancy for the first example.

**Step 6:** Complete the process. Finally, we update the network N with simplified crosstalk friendly Boolean expression which is

N=$X_{or}(X_{hetero}(X_{and}(A_3,B_3'),B_0',B_2),X_{hetero}(X_{homo}(X_{nand}(A_1',B_1),A_2, B_2'),A_3,B_3'))$

**Second example:**

**Step 1:** Input an arbitrary network

In 2-bit multiplier, there are four outputs and four inputs to the network.

$Y_0 = A_0B_0$
$Y_1 = A_1A_0'B_0 + A_1B_1'B_0 + A_1'A_0B_1 + A_0B_1B_0'$
$Y_2 = A_1A_0'B_1 + A_1B_1B_0'$
$Y_3 = A_1A_0B_1B_0$

**Step 2:** By using the SIS [10] tool for preprocessing and decompose, we obtain the following expression:

$$Y_3 = Y_0A_1B_1 \quad (1)$$
$$Y_2 = f_1B_1 \quad (2)$$
$$Y_1 = f_2 A_1 + f_3 \quad (3)$$
$$Y_0 = A_0B_0 \quad (4)$$
$$f_1 = A_0'A_1 + A_1B_0' \quad (5)$$
$$f_2 = A_0'B0 + B_0B_1' + A_0'B_1 \quad (6)$$
$$f_3 = A_0 A_1'B_1 \quad (7)$$

**Step 3** For each function of network N, presented in equation (1)-(7), we have to check if the function is already in Crosstalk homogeneous or heterogeneous form.

- First function $f_1$ is neither in homogeneous nor in heterogeneous form. Next, we checked to see if the function has any common literals. We found that $A_1$ is the common literal in both of the product terms in function $f_1$. Therefore, we factor out the common term and update the function as $f_1 = (A_0'+B_0') A_1$ which is in heterogeneous crosstalk circuit ((A+B)C) form. Next, we proceed to simplify other functions.
- Function $f_2$ is directly in crosstalk homogeneous form (ab+bc+ca), therefore we update the function with crosstalk homogeneous gate and check whether all the functions are simplified or not. As there is another function to be simplified, we go back to synthesize $f_3$.
- As function $f_3$ is also directly in crosstalk homogeneous AND form (ABC) and there is no other function left to be simplified, we move on to step 4.

**Step 4:** Update the node function for inverted output. Since there is no other function to simplify, we check if there exist any function in inverted form. We have found no function to be in inverted form.

**Step 5:** Check for redundant functions and also redundant input to single functions. We have checked and found no redundancy in the simplified network.

**Step 6:** Complete the process. Finally, we update the network N with simplified crosstalk friendly Boolean expression.

$$Y_3 = X_{and}(Y_0,A_1,B_1) \quad (1)$$
$$Y_2 = X_{and}(X_{hetero}(A_1,A_0',B_0'),B_1) \quad (2)$$
$$Y_1 = X_{hetero}(X_{and}(A_0,A_1',B_1),X_{homo}(A_0',B_0,B_1'),A_1) \quad (3)$$
$$Y_0 = X_{and}(A_0,B_0) \quad (4)$$
$$f_1 = A_0'A_1 + A_1B_0' \quad (5)$$
$$f_2 = A_0'B0 + B_0B_1' + A_0'B_1' \quad (6)$$
$$f_3 = A_0 A_1'B_1 \quad (7)$$

IV. COMPARISON RESULTS

Comparison between the proposed approach and previous majority based synthesis approaches [6]-[8] is presented in this section. We have simplified different functions, arithmetic blocks and also three MCNC benchmarks [11]. Table I lists the results for benchmarks. For CMOS, all the primitive cells are considered and for majority based approach, primitive cells are replaced with equivalent Crosstalk gates. For gate count comparison, the inverter is included wherever needed for all three different cases. Our results show significant improvement in a number of gates and transistor count with respect to CMOS. The average reduction (R%) in gate count with respect to CMOS approach is 44%, with the maximum reduction being 77%. For MCNC benchmarks, the average gate count reduction is 44%, with the maximum reduction being 58%. This mostly due to traditional logic reduction approaches for CMOS are constrained to use a limited set of standard cell functions, where, more complex logic functions are not implemented because of the performance concerns that arise in CMOS logic circuits as they would require long pull-up and pull-down branches of switch(transistor) patterns. We also compared our results with majority based simplification approaches due to the similarity between logic reduction approaches. The average reduction (R%) in gate count with respect to other majority synthesis approach is 33%, with the maximum reduction being 50%. For MCNC benchmarks, the average gate count reduction is 20%, with the maximum reduction being 36%. This is mostly due to majority logic approaches are inefficient in logic reduction as they provide a very limited number of primitive gates (majority-three, majority-five, and inverter) and any logic function needs to be transformed to these gates. However, for all the cases, the Crosstalk computing provides holistic logic-reduction opportunities owing to its ability to effectively implement all three, traditional standard cell functions, majority-logic gates, and additional complex functions.

## V. CONCLUSION

We presented a logic simplification approach for large scale Crosstalk circuit integration. We simplified different Boolean networks like complex logic networks obtained from 4-bit ALU, Multiplier, Adder and also three MCNC benchmark circuits. Our results show significant density benefits over CMOS and majority based approach; for the best case, there is 58% and 36% reduction in density over CMOS based and Majority based logic reduction approach respectively. The logic simplification approach presented is a vital step towards full-scale synthesis of Crosstalk circuits leveraging existing EDA tools.

## REFERENCES

1. Naveen kumar Macha, et al., "A New Concept for Computing Using Interconnect Crosstalks," 2017 IEEE International Conference on Rebooting Computing (ICRC), Washington, DC, USA, December 2017.
2. Naveen kumar Macha, Sandeep Geedipally, Bhavana Tejaswee Repalle, Md Arif Iqbal, Wafi Danesh, Mostafizur Rahman "Crosstalk based fine-grained Reconfiguration Techniques for Polymorphic Circuits," IEEE/ACM NANOARCH 2018.
3. Naveen kumar Macha, Bhavana Tejaswini Repalle, Sandeep Geedipally, Rafael Rios, Mostafizur Rahman "A New Paradigm for Fault-Tolerant Computing with Interconnect Crosstalks," 2018 IEEE International Conference on Rebooting Computing (ICRC), Washington, DC, USA, December 2018.
4. Naveen kumar Macha, Sandeep Geedipally, Bhavana Tejaswee Repalle, Md Arif Iqbal, Wafi Danesh, Mostafizur Rahman "A New Paradigm for Computing for Digital Electronics under Extreme Environments," IEEE Aerospace Conference 2019.
5. Rajanikanth Desh, Naveen Kumar Macha, Sehtab Hossain, Repalle Bhavana Tejaswini, Mostafizur Rahman," A Novel Analog to Digital Conversion Concept with Crosstalk Computing," IEEE/ACM International Symposium on Nanoscale Architectures,2018.
6. Zhang, Rui, Pallav Gupta, and Niraj K. Jha. "Synthesis of the majority and minority networks and its applications to QCA, TPL and SET based nanotechnologies." 18th International Conference on VLSI Design held jointly with 4th International Conference on Embedded Systems Design. IEEE, 2005.
7. Kong, Kun, Yun Shang, and Ruqian Lu. "An optimized majority logic synthesis methodology for quantum-dot cellular automata." IEEE Transactions on Nanotechnology 9.2 (2010): 170-183.
8. Wang, Peng, et al. "Comprehensive majority/minority logic synthesis method." 2013 13th IEEE International Conference on Nanotechnology (IEEE-NANO 2013). IEEE, 2013.
9. Cadence Conformal LEC, http://www.cadence.com/products/
10. Sentovich, Ellen M., et al. "SIS: A system for sequential circuit synthesis." (1992)
11. Yang, Saeyang. Logic synthesis and optimization benchmarks user guide: version 3.0. Microelectronics Center of North Carolina (MCNC), 1991.